\documentclass[preprint,showpacs,showkewords,preprintnumber,amsmath,amssymb]{revtex4}
 \usepackage{tipa}
 \usepackage{amsmath}
 \usepackage{txfonts}
 \usepackage{amssymb}
 \usepackage{graphicx,epsfig}
 \usepackage{bm}
 \begin{document}

 \title{Lepton mixing patterns from $PSL_2(7)$ with a generalised CP symmetry  }
 \author{Shu-Jun Rong}\email{rongshj@glut.edu.cn}

 \affiliation{College of Science, Guilin University of Technology, Guilin, Guangxi 541004, China}

 \begin{abstract}
Lepton mixing patterns from the modular group $PSL_2(7)$ with generalised CP symmetries are studied. The residual symmetries in both charged leptons and neutrinos sector are $Z_{2}\times CP$.
Seven types of mixing patterns at the $3\sigma$ level of the new global fit data are obtained. Among these patterns, three types of patterns can give the Dirac CP phase which is in the $1\sigma$ range of the global fit data. The effective mass of neutrinoless double-beta decay for these patterns are also examined.

\end{abstract}


 \maketitle

 \section{Introduction}
CP violation in the hadron sector was observed in 1964\cite{1}. Whether there is a counterpart in the lepton sector is still a mystery. Recent neutrino oscillations experiments show that the 1-3 mixing angle of leptons is nonzero\cite{2,3,4,5}. It intrigues experiments to detect the Dirac CP violating phase.  Specially, some fit results\cite{6,7} hint that this phase is around $-\frac{\pi}{2}$. In the theoretical respect, how to predict nontrivial lepton CP phases is interesting.  In order to obtain mixing parameters of leptons, discrete flavor symmetries are widely used\cite{8,9,10,11,12,13,14,15,16,17,18,19,20,21,22,23,24,25,26,27,28,29,30,31,32,33,34,35,36,37,38,39,40,41}. However, if no perturbation is considered, only finite groups of large orders could accommodate the results of new experiments\cite{35}. Furthermore, they give a trivial Dirac CP violating phase\cite{35}. In order to improve predictions of flavor groups, some efforts have been made in generalizations of symmetries\cite{42,43}. Specially, a intriguing method called generalised CP (GCP) symmetry was introduced\cite{44,45,46,47,48,49,50,51,52,53,54,55,56,57,58,59,60,61,62,63,64,70,71}. In this scenario, the leptonic Largrangian satisfies both flavor and GCP symmetries. After spontaneous symmetries breaking, the residual flavor and GCP  symmetries constrain the structures of mass matrices of leptons. Then information on leptonic mixing mixing angles and CP phases are obtained. From groups $S_{4}$, $A_{5}$ with GCP symmetries, a trivial or maximal Dirac CP phase is obtained\cite{46,56}. The maximal Dirac phase satisfies the $1\sigma$ constraint from the new recent global fit data in case of inverted mass ordering\cite{65}. However, it is not in the $1\sigma$ range for the normal mass ordering. $S_{4}$, $A_{5}$ are small modular groups. We want to know whether a large one could give a more fit CP phase.

In this paper, we study the predictions of the modular group $PSL_2(7)$ with the GCP symmetry in the case of  Majorana neutrinos. We suppose that residual symmetries in neutrinos and charged leptons sector are both $Z_{2}\times CP$. Here $CP$ denotes a GCP symmetry. After examinations of combinations of residual symmetries, we find seven types of mixing patterns at the $3\sigma$ level of the fit data\cite{65}.  Among them, three types satisfy the $1\sigma$ constraint. So the group $PSL_2(7)$ with the GCP symmetry may serve as a candidate for explanations to experiments accommodable mixing patterns. We note that lepton mixing patterns from large finite modular group have been studied in the recent Refs. \cite{66,72,73}. In Refs. \cite{66,72}, no GCP symmetry is considered. In Ref.\cite{72}, residual GCP symmetries are considered either in the neutrino sector or the charged lepton sector. Namely, there is only one unfixed parameter in the lepton mixing matrix.  Here we consider the case that residual GCP symmetries constrain both charged leptons and neutrinos. So two parameters are contained in our mixing patterns.

This paper is organised as follows. In Section II, the framework for the application of the group $PSL_2(7)$ with the GCP symmetry is introduced. In Section III, the results from examination of the residual symmetries are presented. Finally, a summary is made.

\section{Framework}
In this section, we describe the basic facts of the group  $PSL_2(7)$ and introduce the method of deriving lepton mixing patterns from the residual flavor and GCP symmetries.

\subsection{Group theory of $PSL_2(7)$ }
\subsubsection{Generic facts}
The group $PSL_2(7)$ is also named $\Sigma(168)$. It could be constructed with two generators which satisfy following relations\cite{66}:
\begin{equation}
\label{eq:1}
S^{2}=T^{7}=E,~~(ST)^{3}=(ST^{-1}ST)^{4}=E,
\end{equation}
where $E$ is the identity element.
This group has 6 conjugacy classes listed as follows\cite{66}:
\begin{equation}
\label{eq:2}
1\mathcal{C}_{1}: E, ~~21\mathcal{C}_{2}: S, ~~56\mathcal{C}_{3}:ST,  ~~42\mathcal{C}_{4}:ST^{3}, ~~24\mathcal{C}_{7}^{1}: T, ~~24\mathcal{C}_{7}^{2}: T^{3},
\end{equation}
where $i\mathcal{C}_{j}$ denotes that the class contains $i$ elements of order $j$. Accordingly, there are 6 irreducible representations, namely,
\begin{equation}
\label{eq:3}
\mathbf{1}~, ~ \mathbf{3}~, ~ \mathbf{3}^{*}~, ~\mathbf{6}~, ~\mathbf{7}~, ~\mathbf{8}~.
\end{equation}
Without loss of generality, we consider the 3-dimensional representation $ \mathbf{3}$ in following sections. Accordingly, the generators could be expressed as\cite{66}
\begin{equation}
\label{eq:4}
S=\frac{2}{\sqrt{7}}\left(
  \begin{array}{ccc}
   s_{1} & s_{2} & s_{3} \\
    s_{2} & -s_{3} & s_{1} \\
    s_{3} & s_{1} & -s_{2} \\
  \end{array}
\right),~~T=
\left(
  \begin{array}{ccc}
   \varphi_{7}^{2}& 0 & 0 \\
    0 &\varphi_{7} & 0 \\
    0 & 0 & \varphi_{7}^{*3} \\
  \end{array}
\right),
\end{equation}
where $s_{k}=\sin{\frac{k\pi}{7}}$, $\varphi_{7}=e^{i 2\pi/7}$.

 Resorting to the conjugacy classes, we can obtain abelian subgroups of
$PSL_2(7)$. These groups are candidates of the residual symmetries for leptons. In this paper we consider the residual symmetry $Z_{2}\times CP$ for leptons. So $Z_{2}$ subgroups are relevant. There are 21 $Z_{2}$ subgroups which are identified with the generators of them\cite{66}, i.e.,
\begin{equation}
\label{eq:5}
\begin{array}{c}
A_{1}:S,~A_{2}:T^{2}ST^{3}ST,~A_{3}:TST^{3}ST^{2},~A_{4}:T^{4}ST^{3},~A_{5}:T^{3}ST^{4},~A_{6}:T^{2}ST^{4}ST^{2},\\
A_{7}:ST^{2}ST^{4}ST^{2}S, ~A_{8}:ST^{4}ST^{3}S,~A_{9}:ST^{3}ST^{4}S,~A_{10}:T^{5}ST^{2},~A_{11}:T^{2}ST^{5},\\
A_{12}:T^{6}ST,~A_{13}:TST^{6},~A_{14}:ST^{4}ST^{4},~A_{15}:ST^{3}ST^{3},A_{16}:ST^{2}ST,~A_{17}:ST^{5}ST^{6},\\
A_{18}:(T^{2}ST^{3}S)^{2},~A_{19}:(T^{5}ST^{4}S)^{2},~A_{20}:(ST^{3}ST^{4})^{2},~A_{21}:(ST^{4}ST^{3})^{2}.
\end{array}
\end{equation}

\subsubsection{Automorphism of $PSL_2(7)$}
An automorphism of a group is a transformation which permutates elements of the group. These transformations form a group, namely the automorphism group. For the group $PSL_2(7)$, the structure of the automorphism group is simple. It is listed as follows:
\begin{equation}
\label{eq:6}
\begin{array}{c}
\mathrm{Z}(PSL_2(7))=Z_{1}, ~~\mathrm{Aut}(PSL_2(7))\cong PSL(2,Z_{7})
\rtimes Z_{2},\\
\mathrm{Inn}(PSL_2(7))\cong PSL_2(7),~~\mathrm{Out}(PSL_2(7))\cong Z_{2} = \{id, u\},
\end{array}
\end{equation}
where Z, Aut, Inn, Out denote the centre, the automorphism group, the inner automorphism, and the outer automorphism group respectively.
In detail, the inner automorphism group is composed of permutations of elements in the same conjugacy class. The outer
automorphism group swaps conjugacy classes and representations. So it reflects the symmetries of the character table shown in Table~\ref{tab:1}.
The unique nontrivial outer automorphism of the group $PSL_2(7)$ is
\begin{equation}
\label{eq:7}
u:~24\mathcal{C}_{7}^{1}\leftrightarrow24\mathcal{C}_{7}^{2},~~ \mathbf{3}\leftrightarrow\mathbf{3}^{*}.
\end{equation}
The representation of $u$ could be obtained from its action on the generators $S$, $T$, i.e.,
\begin{equation}
\label{eq:8}
u:~S\leftrightarrow S,~~ T\leftrightarrow T^{*}=T^{6}.
\end{equation}
In the 3-dimensional representation, the specific equations of the transformation read
\begin{equation}
\label{eq:9}
X(u)S^{*} X^{-1}(u)=S^{-1}=S,~ X(u)T^{*} X^{-1}(u)=T^{-1}=T^{*}.
\end{equation}
The solution is
\begin{equation}
\label{eq:10}
X(u)=e^{i\alpha}diag (1,~1,~1).
\end{equation}
Since the global phase is trivial for the lepton mixing patterns, we choose $e^{i\alpha}=1$ in the following sections. A general automorphism is the product of the inner and the outer one. It could be expressed  as
\begin{equation}
\label{eq:11}
X(g_{i})=\rho_{3}(g_{i})X(u)=\rho_{3}(g_{i}),~~with~ g_{i}\in PSL_2(7),
\end{equation}
where $\rho_{3}(g_{i})$ is the 3-dimensional representation of the group element.

\begin{table}
\caption{ Character table of the group $PSL_2(7)$ \cite{66}}
\label{tab:1}       
\begin{tabular}{||c|c|c|c|c|c|c||}
\noalign{\smallskip}\hline
~~Rep.~~ &~~~~~~$1C_{1}$~~~~~~& ~~~~~~$21C_{2}$~~~~~~&~~~~~~$56C_{3}$~~~~~~ & ~~~~~~$42C_{4}$~~~~~~ & ~~$24C_{7}^{1}$~~& ~~$24C_{7}^{2}$~~\\[0.5ex]\hline
\noalign{\smallskip}\noalign{\smallskip}\hline
~~$\mathbf{1}$~~ & ~~1~~ &~~1~~ & ~~1~~&~~1~~ &~1~~ &~~1~~  \\\hline
~~$\mathbf{3}$~~&~~3~~ & ~~ -1~~ & ~~0~~&~~1~~&~$\varphi_{7}^{*}+\varphi_{7}^{*2}+\varphi_{7}^{*4}$~~ &~~$\varphi_{7}+\varphi_{7}^{2}+\varphi_{7}^{4}$~~  \\\hline
~~$\mathbf{3}^{*}$~~&~~3~~ &~~-1~~ & ~~0~~&~~ 1~~ &~$\varphi_{7}+\varphi_{7}^{2}+\varphi_{7}^{4}$~~ &~~$\varphi_{7}^{*}+\varphi_{7}^{*2}+\varphi_{7}^{*4}$~~  \\\hline
~~$\mathbf{6}$~~& ~~6~~ & ~~2~~ & ~~0~~&~0~~ &~-1~~ &~~-1~~ \\\hline
~~$\mathbf{7}$~~~& ~~7~~ & ~~-1~~ & ~~1~~&~-1~~&~0~~ &~~0~~   \\\hline
~~$\mathbf{8}$~~& ~~8~~ & ~~0~~ & ~~-1~~&~~0~~&~1~~ &~~1~~  \\\hline
\noalign{\smallskip}
\end{tabular}
\vspace*{0.5cm}  
\end{table}

\subsection {Approach}

\subsubsection{GCP compatible with $PSL_2(7)$}

The GCP transformation acts on the flavor space as
\begin{equation}
\label{eq:12}
\Phi\rightarrow X\Phi^{C},
\end{equation}
where $\Phi$ is a multiplet of fields, $X$ is a unitary matrix, and $\Phi^{C}$ is the CP conjugation of $\Phi$.
In contrast, the flavor group acts on the fields as
\begin{equation}
\label{eq:13}
\Phi\rightarrow \rho(g_{i})\Phi,~~with ~~g_{i}\in PSL_{2}(7).
\end{equation}
Accordingly, the consistence condition of GCP is\cite{45}
\begin{equation}
\label{eq:14}
(X^{-1}\rho(g)X)^{*}=\rho(g^{\prime}).
\end{equation}
Therefore, $X$ is an automorphism of the flavor group $PSL_{2}(7)$. These GCP transformations form an automorphism group $CP$. The general theory satisfies the symmetry
$PSL_{2}(7)\rtimes CP$. After fermions obtain masses from the vacuum expectation values of scalar fields, the original symmetry $PSL_{2}(7)\rtimes CP$ is broken to $G_{e}\rtimes CP_{e}$ in the charged lepton sector and $G_{\nu}\rtimes CP_{\nu}$ in the neutrino sector. Thus the mass matrices of charged leptons and Majorana neutrinos satisfy the relations
\begin{equation}
\label{eq:15}
\rho^{+}(g_{e})m_{e}m_{e}^{+}\rho(g_{e})=m_{e}m_{e}^{+},~~with~~ g_{e}\in G_{e},
\end{equation}
\begin{equation}
\label{eq:16}
\rho^{T}(g_{\nu})m_{\nu}\rho(g_{\nu})=m_{\nu},~~with~~ g_{\nu}\in G_{\nu}.
\end{equation}
The CP transformation $X$ follows the relations
\begin{equation}
\label{eq:17}
~~X^{+}_{e}m_{e}m_{e}^{+}X_{e}=(m_{e}m_{e}^{+})^{*},~~ (X^{-1}_{e}\rho(g_{e})X_{e})^{*}=\rho(g^{\prime}_{e}).
\end{equation}
\begin{equation}
\label{eq:18}
~~X^{T}_{\nu}m_{\nu}X_{\nu}=m^{*}_{\nu},~~ (X^{-1}_{\nu}\rho(g_{\nu})X_{\nu})^{*}=\rho(g^{\prime}_{\nu}).
\end{equation}
Since masses of leptons are non-degenerate, the $CP$ transformation $X$ should be a symmetric unitary matrix~\cite{56}, i.e.,
\begin{equation}
\label{eq:19}
X_{\alpha}=X^{T}_{\alpha},~~X_{\alpha}X^{*}_{\alpha}=E, ~~with~~\alpha=e,\nu~(no ~sum).
\end{equation}
So $X_{\alpha}$ can be decomposed as $X_{\alpha}=\Omega_{\alpha}\Omega_{\alpha}^{T}$.
This kind of $CP$ transformations is called Bickerstaff-Damhus automorphism (BDA)~\cite{67,68}. For the group $PSL_{2}(7)$, all BDAs in the 3-dimensional representation are listed as follows:
\begin{equation}
\label{eq:20}
\begin{array}{c}
  E,~T^{i},~ST^{i}S, ~~with ~~i=1,2,3,4,5,6, \\
 T^{3}ST^{3}, ~TST^{4}ST,~TST^{5}ST,~ST^{3}ST^{3}S,~STST^{4}STS,~STST^{5}STS,\\
 (T^{2}ST^{2})^{j},(ST^{2}ST^{2}S)^{j}, (T^{2}ST^{5}ST^{2})^{j},~~with ~~j=1,2,3.
\end{array}
\end{equation}

\subsubsection{Mixing patterns from residual symmetries $Z_{2}\times CP$}
Once the residual symmetries are fixed, the lepton mixing pattern could be obtained up to permutations of rows and columns. In the direct method, the mixing matrix is completely determined by the symmetries. In the semidirect method, only several elements of the matrix are certain because of degeneracy of the eigenvalues of the residual symmetries. We concern on the semidirect method in this paper. The residual symmetry is $Z_{2e}\times CP_{e}$, $Z_{2\nu}\times CP_{\nu}$ in charged leptons and neutrinos sector respectively. The consistence equation is written as
\begin{equation}
\label{eq:21}
X\rho^{*}(g_{e,\nu})X^{*}=\rho(g_{e,\nu}), ~~with~g_{e,\nu}\in Z_{2}.
\end{equation}
Accordingly, the lepton mixing matrix $U_{PMNS}\equiv U^{+}_{e}U_{\nu}$ is obtained from the matrix \cite{46}
\begin{equation}
\label{eq:22}
U_{e,\nu}=\Omega_{e,\nu} O(\theta_{e,\nu})P_{e, \nu},
\end{equation}
where $R_{e,\nu}$ is a rotation matrix with an angle parameter $\theta_{e,\nu}$, $P_{e,\nu}$ is a phase matrix, i.e.,
\begin{equation}
\label{eq:23}
P_{e,\nu}=diag(1,~ i^{j}, ~i^{k}), ~~with ~~j, k=0, 1, 2, 3.
\end{equation}
Because $P_{e}$ gives nonphysical phases, it is omitted in the following sections.

\subsubsection{Similar transformations}
In order to obtain viable mixing patterns, all possible combinations of residual symmetries should be examined. However, if two combinations are connected by a similarity transformation, namely
\begin{equation}
\label{eq:24}
\rho(g^{\prime}_{e,\nu})=V\rho(g_{e,\nu})V^{+},~~
X^{\prime}_{e,\nu}=VX_{e,\nu}V^{T},
\end{equation}
they would correspond to the same mixing matrix.
Therefore, we could just examine nonequivalent combinations. In following sections, $Z_{2\nu}$ is fixed on the subgroup $Z_{2}^{S}$ which is generated by the group element $S$. The consistent GCP transformations for $Z^{S}_{2}$ are listed as follows
\begin{equation}
\label{eq:25}
X_{1}=E,~~X_{2}=S,~~X_{3}=T^{2}ST^{5}ST^{2},~~ X_{4}=(T^{5}ST^{2}ST^{5})=(T^{2}ST^{5}ST^{2})^{*},
\end{equation}
where $X_{1}$ and $X_{2}$ correspond to the equivalent mixing patterns. So do $X_{3}$ and $X_{4}$.
$Z_{2e}$, $X_{e}$ can be obtained from the similar transformations.
In detail, for generators of $Z_{2}$ subgroups, we have $\rho(A_{i})=V_{i}SV^{+}_{i}$ with $V_{i}$ listed as follows:
\begin{equation}
\label{eq:26}
\nonumber
  V_{2}=\left(
     \begin{array}{ccc}
       0 & 0 & 1 \\
       \varphi^{3}_{7} & 0 & 0 \\
       0 & -\varphi_{7} & 0 \\
     \end{array}
   \right),~~V_{3}=\left(
     \begin{array}{ccc}
       0 & 0 & 1 \\
       \varphi^{*3}_{7} & 0 & 0 \\
       0 & -\varphi^{*}_{7} & 0 \\
     \end{array}
   \right),~~V_{4}=\left(
     \begin{array}{ccc}
       1 & 0 & 0 \\
      0 &  \varphi^{3}_{7} & 0 \\
       0 &0 &  -\varphi_{7} \\
     \end{array}
   \right),
V_{5}=\left(
     \begin{array}{ccc}
       1 & 0 & 0 \\
     0 &  \varphi^{*3}_{7} & 0 \\
       0 &0 &  -\varphi^{*}_{7} \\
     \end{array}
   \right),
\end{equation}
\begin{equation}
\nonumber
  V_{6}=\left(
     \begin{array}{ccc}
       0 & 0 & -1 \\
       -1 & 0 & 0 \\
       0 & 1 & 0 \\
     \end{array}
   \right),~~V_{7}=\left(
     \begin{array}{ccc}
       0 & -1 & 0 \\
       0 & 0 & 1 \\
       -1 & 0 & 0 \\
     \end{array}
   \right),~~V_{8}=\left(
     \begin{array}{ccc}
       0 & 0 & 1 \\
       \varphi^{*}_{7} & 0 & 0 \\
       0 & -\varphi^{2}_{7} & 0 \\
     \end{array}
   \right),~~
   V_{9}=\left(
     \begin{array}{ccc}
       0 & 0 & 1 \\
       \varphi_{7} & 0 & 0 \\
       0 & -\varphi^{*2}_{7} & 0 \\
     \end{array}
   \right),
 \end{equation}
\begin{equation}
~~~~~~~V_{10}=\left(
     \begin{array}{ccc}
       1 & 0 & 0 \\
       0 & \varphi^{2}_{7} & 0 \\
       0 & 0 &  \varphi^{3}_{7}  \\
     \end{array}
   \right),~~ V_{11}=\left(
     \begin{array}{ccc}
       1 & 0 & 0 \\
       0 & \varphi^{*2}_{7} & 0 \\
       0 & 0 &  \varphi^{*3}_{7}  \\
     \end{array}
   \right),~~V_{12}=\left(
     \begin{array}{ccc}
       1 & 0 & 0 \\
       0 & \varphi_{7} & 0 \\
       0 & 0 &  \varphi^{*2}_{7}  \\
     \end{array}
   \right),~~
V_{13}=\left(
     \begin{array}{ccc}
       1 & 0 & 0 \\
       0 & \varphi^{*}_{7} & 0 \\
       0 & 0 &  \varphi^{2}_{7}  \\
     \end{array}
   \right),
 \end{equation}
\begin{equation}
\nonumber
  V_{14}=\left(
     \begin{array}{ccc}
       0 & 1 & 0 \\
       0 & 0 & -\varphi^{3}_{7} \\
      \varphi_{7} & 0 &  0  \\
     \end{array}
   \right),~~ V_{15}=\left(
     \begin{array}{ccc}
       0 & 1 & 0 \\
       0 & 0 & -\varphi^{*3}_{7} \\
      \varphi^{*}_{7} & 0 &  0  \\
     \end{array}
   \right),~~V_{16}=\left(
     \begin{array}{ccc}
       0 & 0 & 1 \\
       \varphi^{2}_{7} & 0 & 0 \\
      0 &-\varphi^{3}_{7} &  0  \\
     \end{array}
   \right),~~
V_{17}=\left(
     \begin{array}{ccc}
       0 & 0 & 1 \\
       \varphi^{*2}_{7} & 0 & 0 \\
      0 &-\varphi^{*3}_{7} &  0  \\
     \end{array}
   \right),
  \end{equation}
\begin{equation}
\nonumber
    V_{18}=\left(
     \begin{array}{ccc}
       0 & 1 & 0 \\
       0 & 0 & -\varphi^{*}_{7} \\
      \varphi^{2}_{7} &0 &  0  \\
     \end{array}
   \right),~~ V_{19}=\left(
     \begin{array}{ccc}
       0 & 1 & 0 \\
       0 & 0 & -\varphi_{7} \\
      \varphi^{*2}_{7} &0 &  0  \\
     \end{array}
   \right),~~ V_{20}=\left(
     \begin{array}{ccc}
       0 & 1 & 0 \\
       0 & 0 & -\varphi^{2}_{7} \\
      \varphi^{3}_{7} &0 &  0  \\
     \end{array}
   \right),~~V_{21}=\left(
     \begin{array}{ccc}
       0 & 1 & 0 \\
       0 & 0 & -\varphi^{*2}_{7} \\
      \varphi^{*3}_{7} &0 &  0  \\
     \end{array}
   \right).
\end{equation}
Specially, $V_{1}$ is the identity matrix. So a general combination of the residual symmetries is of the form
\begin{equation}
\label{eq:27}
(Z^{A_{i}}_{2e}, ~ X_{ej}(A_{i}),~Z^{S}_{2\nu}, ~X_{\nu k}(S)),
\end{equation}
with $j,~ k=1, 3$.
The corresponding lepton mixing matrix is written as
\begin{equation}
\label{eq:28}
U_{PMNS}=O^{T}(\theta_{e})\Omega^{+}_{j}V^{+}_{i}\Omega_{k}O(\theta_{\nu})P_{\nu},
\end{equation}
where
\begin{equation}
\label{eq:29}
O(\theta_{e,\nu})=\left(
        \begin{array}{ccc}
        1 & 0 & 0 \\
         0  & \cos\theta_{e,\nu} & \sin\theta_{e,\nu} \\
          0 &  -\sin\theta_{e,\nu} & \cos\theta_{e,\nu} \\
        \end{array}
      \right),
\end{equation}
\begin{equation}
\label{eq:30}
\Omega_{1}=\left(
\begin{array}{ccc}
r_{1} & -\sin\theta_{1} & -\cos\theta_{1}\sin\theta_{2} \\
 r_{2} & 0 & \cos\theta_{2} \\
 r_{3} & \cos\theta_{1} & -\sin\theta_{1}\sin\theta_{2}
\end{array}
\right),
\end{equation}
\begin{equation}
\label{eq:31}
\Omega_{3}=\Omega_{1}\cdot\left(
        \begin{array}{ccc}
        1 & 0 & 0 \\
         0  & e^{i\pi/4}\cos\phi & e^{i3\pi/4}\sin\phi \\
          0 &  -e^{i\pi/4}\sin\phi & e^{i3\pi/4}\cos\phi \\
        \end{array}
      \right),
\end{equation}
with
\begin{equation}
\label{eq:32}
  r_{1}=2\sqrt{\frac{2}{7}}\sin\frac{2\pi}{7}\sin\frac{3\pi}{7},~~r_{2}=2\sqrt{\frac{2}{7}}\sin\frac{\pi}{7}\sin\frac{2\pi}{7},
  ~~r_{3}=2\sqrt{\frac{2}{7}}\sin\frac{\pi}{7}\sin\frac{3\pi}{7},
\end{equation}
\begin{equation}
\label{eq:33}
  \theta_{1}=\arcsin\frac{r_{3}}{\sqrt{r^{2}_{1}+r^{2}_{3}}},~~ \theta_{2}=\arcsin\frac{r_{2}\cos\theta_{1}}{\sqrt{r^{2}_{1}+r^{2}_{2}\cos^{2}\theta_{1}}}, \phi=\arcsin\frac{1}{\sqrt{1+x^{2}_{1}}},
\end{equation}
where $x_{1}$ is a real root of the equation
\begin{equation}
\label{eq:34}
x^{12}-48x^{10}+323x^{8}-608x^{6}+323x^{4}-48x^{2}+1=0,
\end{equation}
$x_{1}\simeq\pm0.449807$.

\section{Results}

\subsection{Viable mixing matrices from combinations of residual symmetries}
Given the recent global fit data of neutrino oscillations\cite{65}, we examine the predictions of combinations of residual symmetries of the form ($Z^{A_{i}}_{2e}$, ~ $X_{ej}(A_{i})$,~$Z^{S}_{2\nu}$, ~$X_{\nu k}(S))$ with the $\chi^{2}$ function defined as
\begin{equation}
\label{eq:35}
\chi^{2}=\sum_{ij=13,23,12}(\frac{\sin^{2}\theta_{ij}-(\sin^{2}\theta_{ij})^{ex}}{\sigma_{ij}})^{2},
\end{equation}
where $(\sin^{2}\theta_{ij})^{ex}$ is the best fit data from Ref~\cite{65}, $\sigma_{ij}$ is the 1$\sigma$ error. The viable combinations at the $3\sigma$ level (up to equivalent ones) are listed as follows.\\
Type Ia:
\begin{equation}
\label{eq:36}
  (Z_{2e}^{ ST^{4}ST^{4}},~ X_{3e}=V_{16}(T^{2}ST^{5}ST^{2})V^{T}_{16}, ~Z_{2\nu}^{ S},~ X_{1\nu}=E),
\end{equation}
Type Ib:
\begin{equation}
\label{eq:37}
  (Z_{2e}^{ST^{4}ST^{4}},~ X_{1e}=V_{16}V^{T}_{16}, ~Z_{2\nu} ^{S},~ X_{3\nu}=T^{2}ST^{5}ST^{2}),
\end{equation}
Type Ib$^{\ast}$:
\begin{equation}
\label{eq:38}
(Z_{2e}^{ST^{3}ST^{3}},~ X_{1e}=V_{17}V^{T}_{17}, ~Z_{2\nu} ^{S},~ X_{3\nu}=T^{2}ST^{5}ST^{2}),
\end{equation}
Type IIa:
\begin{equation}
\label{eq:39}
(Z_{2e}^{T^{4}ST^{3}},~ X_{1e}=V_{4}V^{T}_{4}, ~Z_{2\nu} ^{S},~ X_{1\nu}=E),
\end{equation}
Type IIa$^{\ast}$:
\begin{equation}
\label{eq:40}
(Z_{2e}^{T^{3}ST^{4}},~ X_{1e}=V_{5}V^{T}_{5}, ~Z_{2\nu} ^{S},~ X_{1\nu}=E),
\end{equation}
Type IIb:
\begin{equation}
\label{eq:41}
(Z_{2e} ^{T^{4}ST^{3}},~ X_{1e}=V_{4}V^{T}_{4}, ~Z_{2\nu} ^{S},~ X_{3\nu}=T^{2}ST^{5}ST^{2}),
\end{equation}
Type IIb$^{\ast}$:
\begin{equation}
\label{eq:42}
(Z_{2e} ^{T^{3}ST^{4}},~ X_{1e}=V_{5}V^{T}_{5}, ~Z_{2\nu} ^{S},~ X_{3\nu}=T^{2}ST^{5}ST^{2}).
\end{equation}
The corresponding mixing matrices are dependent on permutations of rows and columns. For every combination, the matrix which fits the data best is listed as follows.
\begin{equation}
\label{eq:43}
U_{Ia}=S_{13}O^{T}(\theta_{e})\Omega^{+}_{3}V^{+}_{16}\Omega_{1}O(\theta_{\nu})P_{\nu} ,
\end{equation}
\begin{equation}
\label{eq:44}
U_{Ib}=S_{12}O^{T}(\theta_{e})\Omega^{+}_{1}V^{+}_{16}\Omega_{3}O(\theta_{\nu})P_{\nu}S_{12} ,
\end{equation}
\begin{equation}
\label{eq:45}
U_{Ib^{\ast}}=U^{\ast}_{Ib}=S_{12}O^{T}(\theta_{e})\Omega^{+}_{1}V^{+}_{17}\Omega^{\ast}_{3}O(\theta_{\nu})P_{\nu}S_{12} ,
\end{equation}
\begin{equation}
\label{eq:46}
U_{IIa}=S_{13}O^{T}(\theta_{e})\Omega^{+}_{1}V^{+}_{4}\Omega_{1}O(\theta_{\nu})P_{\nu} ,
\end{equation}
\begin{equation}
\label{eq:47}
U_{IIa^{\ast}}=U^{\ast}_{IIa}=S_{13}O^{T}(\theta_{e})\Omega^{+}_{1}V^{+}_{5}\Omega_{1}O(\theta_{\nu})P_{\nu},
\end{equation}
\begin{equation}
\label{eq:48}
U_{IIb}=S_{12}O^{T}(\theta_{e})\Omega^{+}_{1}V^{+}_{4}\Omega_{3}O(\theta_{\nu})P_{\nu},
\end{equation}
\begin{equation}
\label{eq:49}
U_{IIb^{\ast}}=U^{\ast}_{IIb}=S_{12}O^{T}(\theta_{e})\Omega^{+}_{1}V^{+}_{5}\Omega^{\ast}_{3}O(\theta_{\nu})P_{\nu},
\end{equation}
where
\begin{equation}
\label{eq:50}
S_{12}=\left(
         \begin{array}{ccc}
           0 & 1 & 0 \\
           1 & 0 & 0 \\
           0 & 0 & 1 \\
         \end{array}
       \right), ~~~S_{13}=\left(
         \begin{array}{ccc}
           0 & 0 & 1 \\
           0 & 1 & 0 \\
           1 & 0 & 0 \\
         \end{array}
       \right).
\end{equation}
We note that the mixing matrices except $U_{Ia}$ are paired through the complex conjugation. The predictions of the matrices in a pair are identical except the signs of the CP phases.  So we can consider $U_{Ia}$, $U_{Ib}$, $U_{IIa}$, $U_{IIb}$ as representatives.

\subsection{Mixing angles and CP invariants}
Employing the parametrization of the form
\begin{equation}
\label{eq:51}
U_{PMNS}=
\left(
\begin{array}{ccc}
 c_{12}c_{13} & s_{12}c_{13} & s_{13}e^{-i\delta} \\
 -s_{12}c_{23}-c_{12}s_{13}s_{23}e^{i\delta} & c_{12}c_{23}-s_{12}s_{13}s_{23}e^{i\delta} & c_{13}s_{23} \\
s_{12}s_{23}-c_{12}s_{13}c_{23}e^{i\delta} & -c_{12}s_{23}-s_{12}s_{13}c_{23}e^{i\delta} & c_{13}c_{23}
\end{array}
\right)
\left(
\begin{array}{ccc}
1 & 0 & 0 \\
 0 & e^{i\alpha/2} & 0 \\
0 & 0 & e^{i(\beta/2+\delta)}
\end{array}
\right),
\end{equation}
where $s_{ij}\equiv\sin{\theta_{ij}}$, $c_{ij}\equiv\cos{\theta_{ij}}$, $\delta$ is the Dirac CP violating phase, $\alpha$ and $\beta$ are Majorana phases,
we could obtain lepton mixing angles and the CP invariants $J_{cp}$\cite{69}, $J_{1}$, $J_{2}$ defined as
\begin{equation}
\label{eq:52}
J_{cp}\equiv\mathrm{Im}[U_{11}U^{*}_{13}U^{*}_{31}U_{33}]=\frac{1}{8}\sin2\theta_{13}\sin2\theta_{23}\sin2\theta_{12}\cos\theta_{13}\sin\delta.
\end{equation}
 \begin{equation}
 \label{eq:53}
J_{1}\equiv\mathrm{Im}[(U_{11}^{*})^{2}U_{12}^{2}]=\sin^{2}\theta_{12}\cos^{2}\theta_{12}\cos^{4}\theta_{13}\sin\alpha,
\end{equation}
\begin{equation}
\label{eq:54}
J_{2}\equiv\mathrm{Im}[(U_{11}^{*})^{2}U_{13}^{2}]=\sin^{2}\theta_{13}\cos^{2}\theta_{13}\cos^{2}\theta_{12}\sin\beta.
\end{equation}
Their specific forms are listed as follows.\\
$U_{Ia}$:
\begin{equation}
\label{eq:55}
\begin{aligned}
\sin^{2}\theta_{13}(\theta_{e},\theta_{\nu})\simeq &0.3125-0.1758\cos2\theta_{\nu}+0.02073\cos(2\theta_{\nu}-2\theta_{e})-0.1654\cos2\theta_{e}\\
 &+0.2786\cos(2\theta_{\nu}+2\theta_{e})+0.065086\sin2\theta_{\nu}-0.02338\sin(2\theta_{\nu}-2\theta_{e})\\
 &-0.08839\sin2\theta_{e}+0.03824\sin(2\theta_{\nu}-2\theta_{e}),
\end{aligned}
\end{equation}
\begin{equation}
\label{eq:56}
\begin{aligned}
\sin^{2}\theta_{23}(\theta_{e},\theta_{\nu})\simeq &[0.002647\cos^{2}\theta_{e}\cos^{2}\theta_{\nu}+0.05023\cos^{2}\theta_{e}\sin2\theta_{\nu} +0.9531\cos^{2}\theta_{e}\sin^{2}\theta_{\nu}\\
&+0.02677\sin2\theta_{e}\cos^{2}\theta_{\nu}+0.2579\sin2\theta_{e}\sin2\theta_{\nu} +0.27067\sin^{2}\theta_{e}\cos^{2}\theta_{\nu} \\
&+0.150\sin2\theta_{e}\sin^{2}\theta_{\nu}+0.07994\sin^{2}\theta_{e}\sin2\theta_{\nu}\\
&+0.02361\sin^{2}\theta_{e}\sin^{2}\theta_{\nu}] /(1-\sin^{2}\theta_{13}(\theta_{e},\theta_{\nu})),
\end{aligned}
\end{equation}
\begin{equation}
\label{eq:57}
\sin^{2}\theta_{12}(\theta_{e},\theta_{\nu})=\frac{\sin^{2}\theta_{23}(\theta_{e}+\pi/2,\theta_{\nu}+\pi/2)(1-\sin^{2}\theta_{13}(\theta_{e}+\pi/2,\theta_{\nu}+\pi/2))}{1-\sin^{2}\theta_{13}(\theta_{e},\theta_{\nu})},
\end{equation}
\begin{equation}
\label{eq:58}
J_{cp}(\theta_{e},\theta_{\nu})=0,~~~~ J_{1}(\theta_{e},\theta_{\nu})=0,~~~~J_{2}(\theta_{e},\theta_{\nu})=0.
\end{equation}

$U_{Ib}$:
\begin{equation}
\label{eq:59}
\begin{aligned}
\sin^{2}\theta_{13}(\theta_{e},\theta_{\nu})\simeq &0.2438\cos^{2}\theta_{e}\cos^{2}\theta_{\nu}+0.1871\cos^{2}\theta_{e}\sin2\theta_{\nu}+0.01618\cos^{2}\theta_{e}\sin^{2}\theta_{\nu}\\
 &+0.264\sin2\theta_{e}\cos^{2}\theta_{\nu}+0.1659\sin2\theta_{e}\sin2\theta_{\nu}-0.16566\sin2\theta_{e}\sin^{2}\theta_{\nu}\\
 &+0.712\sin^{2}\theta_{e}\cos^{2}\theta_{\nu}-0.187\sin^{2}\theta_{e}\sin2\theta_{\nu}+0.13245\sin^{2}\theta_{e}\sin^{2}\theta_{\nu},
\end{aligned}
\end{equation}
\begin{equation}
\label{eq:60}
\sin^{2}\theta_{23}(\theta_{e},\theta_{\nu})\simeq[0.375-0.3307\cos2\theta_{\nu}] /(1-\sin^{2}\theta_{13}(\theta_{e},\theta_{\nu})),
\end{equation}
\begin{equation}
\label{eq:61}
\sin^{2}\theta_{12}(\theta_{e},\theta_{\nu})\simeq[0.375+0.2194\cos2\theta_{e}-0.2475\sin2\theta_{e}] /(1-\sin^{2}\theta_{13}(\theta_{e},\theta_{\nu})),
\end{equation}
\begin{equation}
\label{eq:62}
\begin{aligned}
J_{cp}(\theta_{e},\theta_{\nu})\simeq &0.001543+0.008446\cos2\theta_{\nu}+0.006903\cos4\theta_{\nu}+0.002733\cos(2\theta_{\nu}-4\theta_{e})    \\
&-0.003911\cos(4\theta_{\nu}-2\theta_{e})+0.03405\cos(2\theta_{\nu}-2\theta_{e})+0.000384\cos(4\theta_{\nu}-4\theta_{e})\\
&-0.002896\cos2\theta_{e}+0.001353\cos4\theta_{e}-0.008932\cos(2\theta_{\nu}+2\theta_{e})\\
&-0.002234\cos(4\theta_{\nu}+4\theta_{e})-0.00114\cos(4\theta_{\nu}+2\theta_{e})-0.00323\cos(2\theta_{\nu}+4\theta_{e})\\
&-0.00951\sin2\theta_{\nu}+0.000585\sin4\theta_{\nu}+0.002935\sin(2\theta_{\nu}-4\theta_{e})\\
&-0.000292\sin(4\theta_{\nu}-2\theta_{e})-0.02566\sin(2\theta_{\nu}-2\theta_{e})+0.000304\sin(4\theta_{\nu}-4\theta_{e})\\
&+0.00328\sin2\theta_{e}-0.00591\sin4\theta_{e}-0.000519\sin(2\theta_{\nu}+2\theta_{e})\\
&+0.00715\sin(4\theta_{\nu}+4\theta_{e})-0.00775\sin(4\theta_{\nu}+2\theta_{e})+0.00388\sin(2\theta_{\nu}+4\theta_{e}),
\end{aligned}
\end{equation}
\begin{equation}
\label{eq:63}
\begin{aligned}
J_{1}(\theta_{e},\theta_{\nu})\simeq &\pm[0.05167-0.00391\cos2\theta_{\nu}-0.00304\cos(2\theta_{\nu}-4\theta_{e}) -0.01371\cos(2\theta_{\nu}-2\theta_{e})   \\
&+0.01037\cos2\theta_{e}+0.006185\cos4\theta_{e}+0.04114\cos(2\theta_{\nu}+2\theta_{e})\\
&-0.00491\cos(2\theta_{\nu}+4\theta_{e})+0.02521\sin(2\theta_{\nu}-4\theta_{e})-0.01546\sin(2\theta_{\nu}-2\theta_{e})\\
&-0.1169\sin2\theta_{e}+0.0153\sin4\theta_{e}-0.0464\sin(2\theta_{\nu}+2\theta_{e})-0.0407\sin(2\theta_{\nu}+4\theta_{e})],
\end{aligned}
\end{equation}
\begin{equation}
\label{eq:64}
\begin{aligned}
J_{2}(\theta_{e},\theta_{\nu})\simeq &\pm[0.00726\cos(2\theta_{\nu}-4\theta_{e}) -0.0248\cos(2\theta_{\nu}-2\theta_{e})+0.05787\cos(2\theta_{\nu}+2\theta_{e})   \\
&-0.05079\cos(2\theta_{\nu}+4\theta_{e})-0.07308\sin2\theta_{\nu}-0.000875\sin(2\theta_{\nu}-4\theta_{e})\\
&+0.02199\sin(2\theta_{\nu}-2\theta_{e})+0.0513\sin(2\theta_{\nu}+2\theta_{e})+0.00612\sin(2\theta_{\nu}+4\theta_{e})].
\end{aligned}
\end{equation}

$U_{IIa}$:
\begin{equation}
\label{eq:65}
\begin{aligned}
\sin^{2}\theta_{13}(\theta_{e},\theta_{\nu})\simeq &0.616\cos^{2}\theta_{e}\cos^{2}\theta_{\nu}-0.133\cos^{2}\theta_{e}\sin2\theta_{\nu}+0.0769\cos^{2}\theta_{e}\sin^{2}\theta_{\nu}\\
 &-0.1328\sin2\theta_{e}\cos^{2}\theta_{\nu}+0.01762\sin2\theta_{e}\sin2\theta_{\nu}-0.1322\sin2\theta_{e}\sin^{2}\theta_{\nu}\\
 &+0.0769\sin^{2}\theta_{e}\cos^{2}\theta_{\nu}-0.1322\sin^{2}\theta_{e}\sin2\theta_{\nu}+0.4155\sin^{2}\theta_{e}\sin^{2}\theta_{\nu},
\end{aligned}
\end{equation}
\begin{equation}
\label{eq:66}
\begin{aligned}
\sin^{2}\theta_{23}(\theta_{e},\theta_{\nu})\simeq &[0.2963+0.0502\cos2\theta_{\nu}-0.1185\cos(2\theta_{\nu}-2\theta_{e})-0.0502\cos2\theta_{e} \\
&-0.101\cos(2\theta_{\nu}+2\theta_{e})-0.01325\sin2\theta_{\nu}+0.01325\sin2\theta_{e}\\
&+0.000294\sin(2\theta_{\nu}+2\theta_{e})] /(1-\sin^{2}\theta_{13}),
\end{aligned}
\end{equation}
\begin{equation}
\label{eq:67}
\sin^{2}\theta_{12}(\theta_{e},\theta_{\nu})=\frac{\sin^{2}\theta_{23}(\theta_{e}+\pi/2,\theta_{\nu}+\pi/2)(1-\sin^{2}\theta_{13}(\theta_{e}+\pi/2,\theta_{\nu}+\pi/2))}{1-\sin^{2}\theta_{13}(\theta_{e},\theta_{\nu})},
\end{equation}
\begin{equation}
\label{eq:68}
J_{CP}(\theta_{e},\theta_{\nu})\simeq0.00919\cos(2\theta_{\nu}-2\theta_{e})-0.01601\cos(2\theta_{\nu}+2\theta_{e})+0.06163\sin(2\theta_{\nu}+2\theta_{e}),
\end{equation}
\begin{equation}
\label{eq:69}
\begin{aligned}
J_{1}(\theta_{e},\theta_{\nu})\simeq &\pm[-0.00475-0.0474\cos2\theta_{\nu}+0.0144\cos(2\theta_{\nu}-4\theta_{e})+0.0203\cos(2\theta_{\nu}-2\theta_{e}) \\
&+0.0541\cos2\theta_{e}-0.0461\cos4\theta_{e}+0.0188\cos(2\theta_{\nu}+2\theta_{e})-0.03285\cos(2\theta_{\nu}+4\theta_{e})\\
&+0.002649\sin2\theta_{\nu}-0.0101\sin(2\theta_{\nu}-4\theta_{e})+0.02699\sin(2\theta_{\nu}-2\theta_{e})+0.02349\sin2\theta_{e}\\
&+0.05424\sin4\theta_{e}-0.01937\sin(2\theta_{\nu}+2\theta_{e})-0.04581\sin(2\theta_{\nu}+4\theta_{e})],
\end{aligned}
\end{equation}
\begin{equation}
\label{eq:70}
J_{2}(\theta_{e},\theta_{\nu})=J_{1}(\theta_{e},\theta_{\nu}+\pi/2).
\end{equation}

$U_{IIb}$:
\begin{equation}
\label{eq:71}
\begin{aligned}
\sin^{2}\theta_{13}(\theta_{e},\theta_{\nu})\simeq &0.2963-0.1324\cos2\theta_{\nu}+0.199\cos(2\theta_{\nu}-2\theta_{e})-0.0502\cos2\theta_{e} \\
&-0.05315\cos(2\theta_{\nu}+2\theta_{e})+0.1464\sin2\theta_{\nu}-0.00634\sin(2\theta_{\nu}-2\theta_{e})\\
&+0.1325\sin2\theta_{e}-0.01973\sin(2\theta_{\nu}+2\theta_{e}),
\end{aligned}
\end{equation}
\begin{equation}
\label{eq:72}
\begin{aligned}
\sin^{2}\theta_{23}(\theta_{e},\theta_{\nu})\simeq &[0.4074+0.2648\cos2\theta_{\nu}-0.2927\sin2\theta] /(1-\sin^{2}\theta_{13}),
\end{aligned}
\end{equation}
\begin{equation}
\label{eq:73}
\sin^{2}\theta_{12}(\theta_{e},\theta_{\nu})=\frac{\sin^{2}\theta_{13}(\theta_{e},\theta_{\nu}+\pi/2)}{1-\sin^{2}\theta_{13}(\theta_{e},\theta_{\nu})},
\end{equation}
\begin{equation}
\label{eq:74}
\begin{aligned}
J_{CP}(\theta_{e},\theta_{\nu})\simeq&0.02616\cos(2\theta_{\nu}-2\theta_{e})+0.02447\cos(2\theta_{\nu}+2\theta_{e})\\
&+0.04\sin(2\theta_{\nu}-2\theta_{e})+0.01804\sin(2\theta_{\nu}+2\theta_{e}),
\end{aligned}
\end{equation}
\begin{equation}
\label{eq:75}
\begin{aligned}
J_{1}(\theta_{e},\theta_{\nu})\simeq &\pm[0.01355+0.01357\cos2\theta_{\nu}+0.02594\cos(2\theta_{\nu}-4\theta_{e})+0.00115\cos(2\theta_{\nu}-2\theta_{e}) \\
&-0.02127\cos2\theta_{e}+0.0643\cos4\theta_{e}-0.03256\cos(2\theta_{\nu}+2\theta_{e})+0.0213\cos(2\theta_{\nu}+4\theta_{e})\\
&-0.03385\sin2\theta_{\nu}-0.04857\sin(2\theta_{\nu}-4\theta_{e})-0.02714\sin(2\theta_{\nu}-2\theta_{e})-0.001069\sin2\theta_{e}\\
&+0.03351\sin4\theta_{e}-0.007202\sin(2\theta_{\nu}+2\theta_{e})-0.002316\sin(2\theta_{\nu}+4\theta_{e})],
\end{aligned}
\end{equation}
\begin{equation}
\label{eq:76}
J_{2}(\theta_{e},\theta_{\nu})=J_{1}(\theta_{e},\theta_{\nu}+\pi/2).
\end{equation}
The sign of $J_{1}$, $J_{2}$ is uncertain. It is dependent on the index $j$, $k$ in the diagonal phase matrix $P_{\nu}$.

\subsection{Constraints on $(\theta_{e},\theta_{\nu})$ and best fit data}
\begin{figure}[tbp]
\centering 
\includegraphics[width=0.48\textwidth]{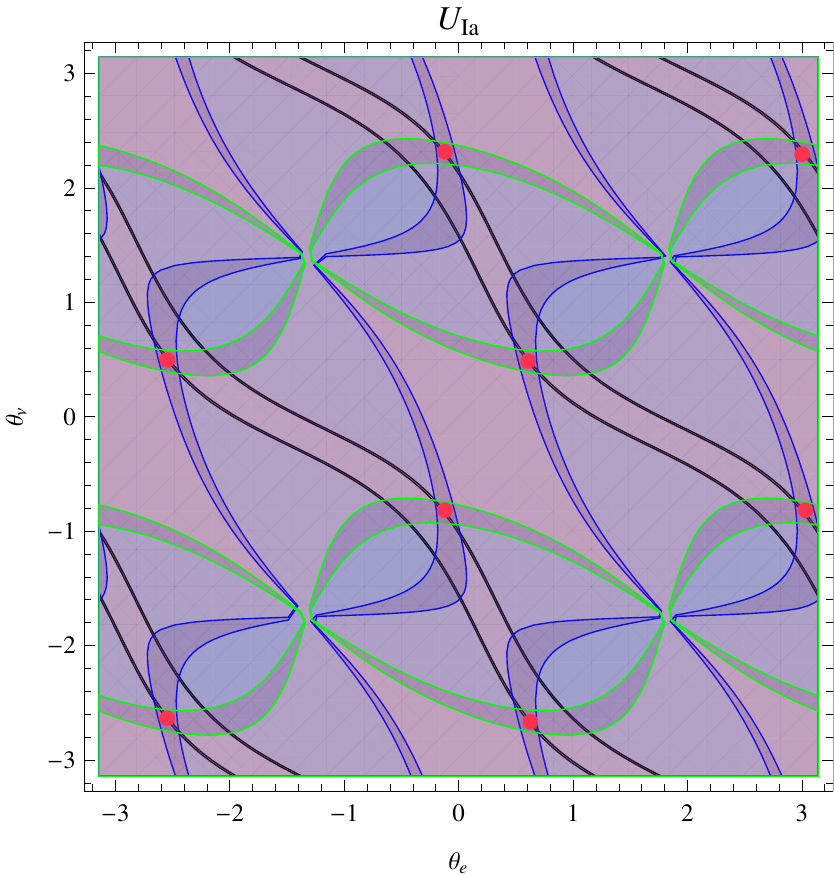}
\hfill
\includegraphics[width=0.48\textwidth]{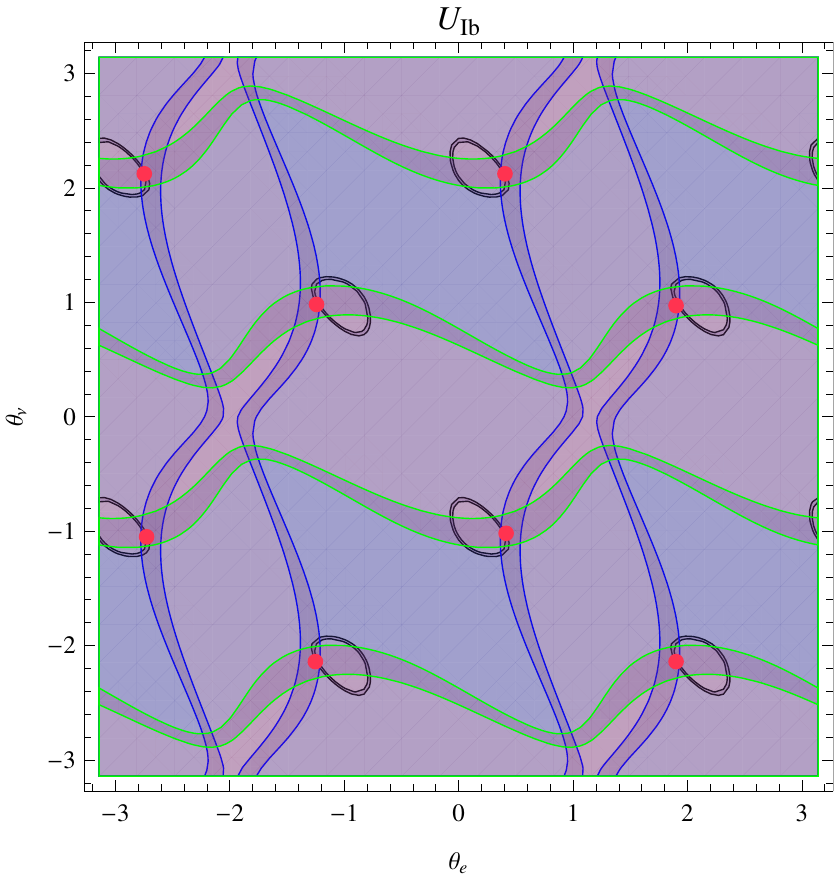}
\hfill
\includegraphics[width=0.48\textwidth]{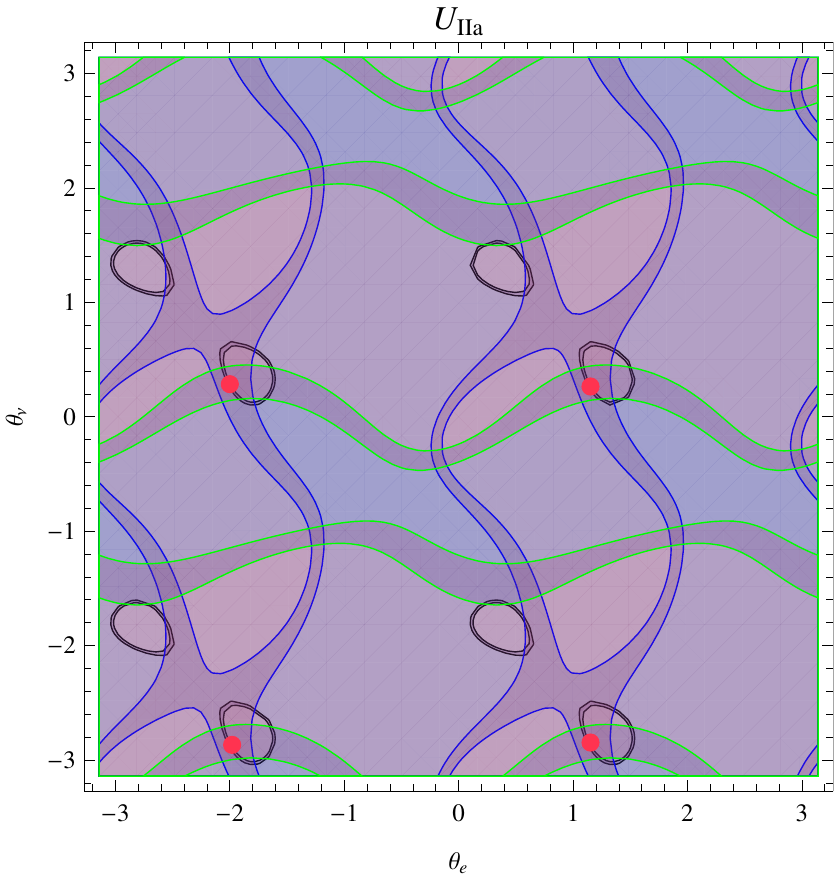}
\hfill
\includegraphics[width=0.48\textwidth]{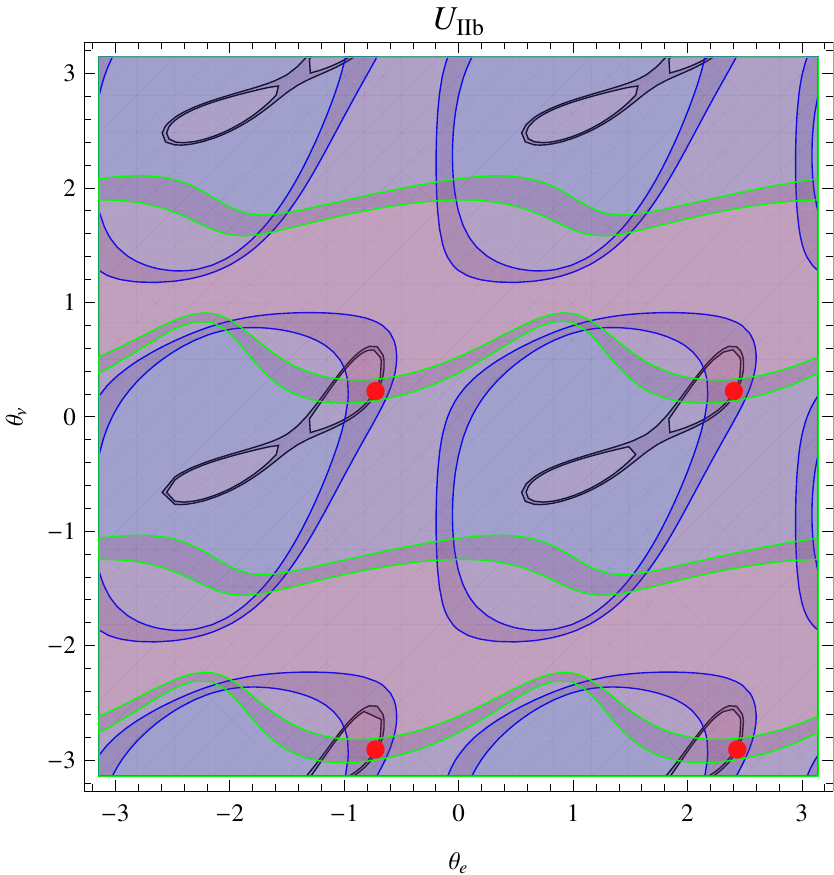}
\hfill
\caption{\label{Fig:1} Parameter-space for the mixing patterns constrained by the global fit data\cite{65} at the $3\sigma$ level in the normal mass ordering(NO). The parameter-spaces for $\theta_{12}$  are the strips with blue boundaries and those for $\theta_{23}$ are with green boundaries. The strips for $\theta_{13}$ are tiny, i.e., almost reduced to black curves. Their intersection areas are signed by the red dots. The parameter-spaces in the case of inverted mass ordering(IO)  are similar. So they are not shown here.}
\end{figure}

The parameter-space of $(\theta_{e},\theta_{\nu})$ is shown in Figure~\ref{Fig:1}. The best fit data of lepton mixing angle and CP phases are listed in Table~\ref{tab:2}. We make some comments on the numerical results:\\
(i). In a period of the parameters $(\theta_{e},\theta_{\nu})$, there are two best fit points for the mixing matrices $U_{Ia}$, $U_{Ib}$. For $U_{Ib}$, these two points give the same magnitude of $\sin\delta$ with different signs. For the mixing matrices $U_{IIa}$, $U_{IIb}$, there is only one best fit point in a period. $\sin\delta$ for them are both positive. Even so, $U^{\ast}_{IIa}$, $U^{\ast}_{IIb}$ could give negative $\sin\delta$ while the mixing angles are kept the same.\\
(ii).~~The best fit value of $\theta_{23}$ from the global fit data\cite{65} is in the second octant. Accordingly, our fit data is in the same octant. In the case of normal mass ordering(NO), the best fit value of $\delta$ in $U_{Ib}$ and $U^{\ast}_{IIb}$ could be around $-0.3\pi$. It is in the $1\sigma$ range of the global fit data. In the case of inverted mass ordering(IO), the best fit value of $\delta$ in $U^{\ast}_{IIa}$ is around $-0.5\pi$. It is also in the $1\sigma$ range of the global fit data.
\begin{table}
\caption{Best fit data of lepton mixing angles and CP phases. 'N' and 'I' denote the normal ordering of neutrino masses and the inverted ordering respectively. }
\label{tab:2}       
\begin{tabular}{||c|c|c|c|c|c|c|c|c|c||}
\noalign{\smallskip}\hline
Patterns &($\theta^{bf}_{e}$,~$\theta^{bf}_{\nu}$)& $\chi^{2}_{min}$&$\sin^{2}\theta_{13}$ & $\sin^{2}\theta_{23}$ & $\sin^{2}\theta_{12}$ & $|\sin\delta|$ & $|\sin\alpha|$& $|\sin\beta|$\\[0.5ex]\hline
\noalign{\smallskip}\noalign{\smallskip}\hline
Ia(N) & (0.18725$\pi$,~0.1644$\pi$),~(-0.0297$\pi$,~0.72267$\pi$)~ &0.000016 & 0.0216 & 0.547&0.320&0&0&0\\
Ia(I) & (0.18596$\pi$,~0.16544$\pi$),~(-0.0297$\pi$,~0.72172$\pi$)~ & 0.021 & 0.0220 & 0.549&0.318&0&0&0\\\hline
Ib(N) &  (-0.4028$\pi$,~0.3393$\pi$),~(-0.8663$\pi$,~-0.3393$\pi)$~&2.33 & 0.0217 & 0.563&0.345&0.813&0.952&0.43\\
Ib(I) &  (-0.4037$\pi$,~0.340$\pi$),~(-0.8654$\pi$,~-0.340$\pi)$ &1.97 & 0.0221 & 0.565&0.343&0.817&0.953&0.42\\\hline
IIa(N) & (0.3694$\pi$,~0.0767$\pi$)~& 0.064 & 0.0216 & 0.549&0.316&0.997&0.769&0.588\\
IIa(I)& (0.3701$\pi$,~0.0745$\pi$)& 0.057 & 0.022 & 0.553&0.316&0.998&0.774&0.547\\\hline
IIb(N) &  (-0.2345$\pi$,~0.06297$\pi$))& 0.79 & 0.02156 & 0.5507&0.304&0.800&0.617&0.747\\
IIb(I) &  (-0.2348$\pi$,~0.06145$\pi$)~& 0.81 & 0.0220 & 0.5546&0.304&0.742&0.616&0.730\\\hline
\noalign{\smallskip}
\end{tabular}
\vspace*{0.5cm}  
\end{table}

\subsection{The effective mass of neutrinoless double-beta decay $<m_{ee}>$}
\begin{figure}[tbp]
\centering 
\includegraphics[width=0.48\textwidth]{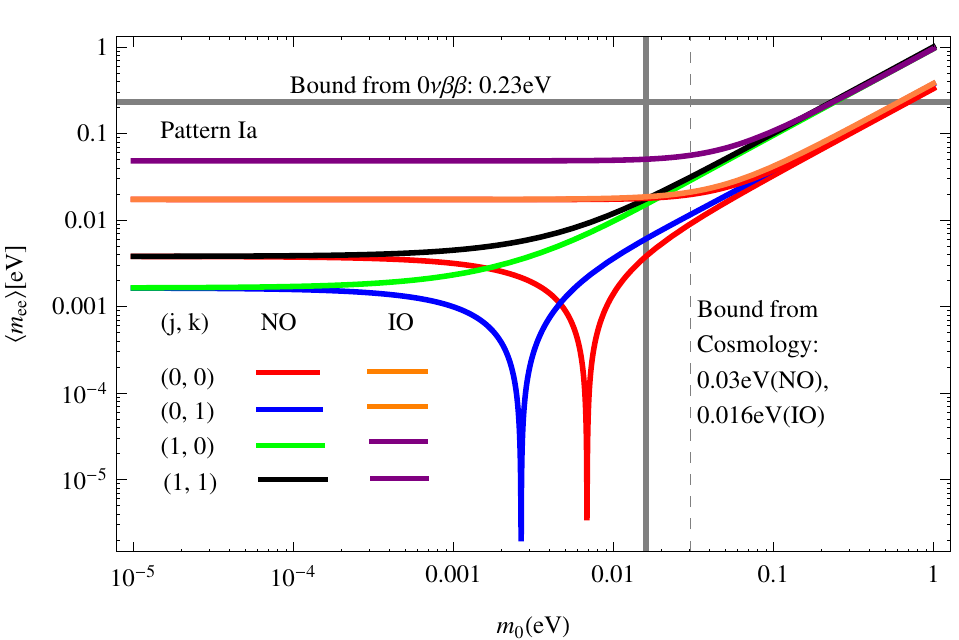}
\hfill
\includegraphics[width=0.48\textwidth]{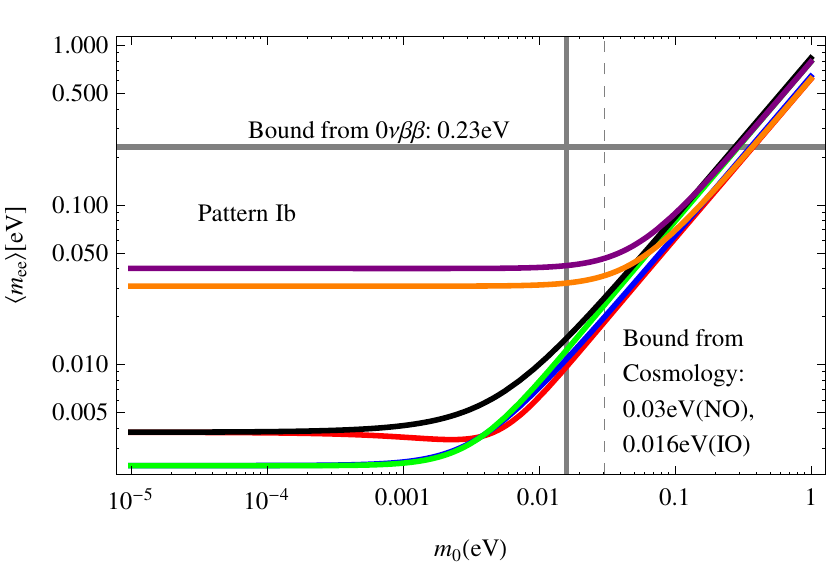}
\hfill
\includegraphics[width=0.48\textwidth]{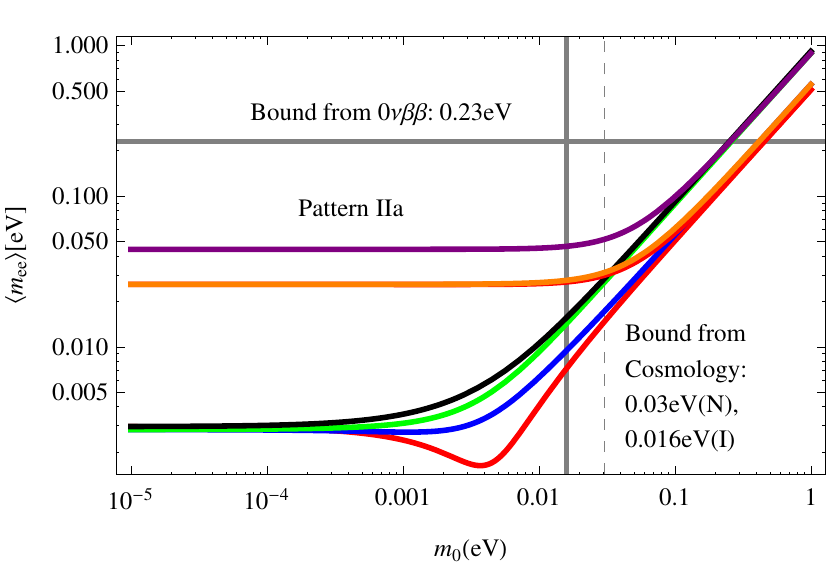}
\hfill
\includegraphics[width=0.48\textwidth]{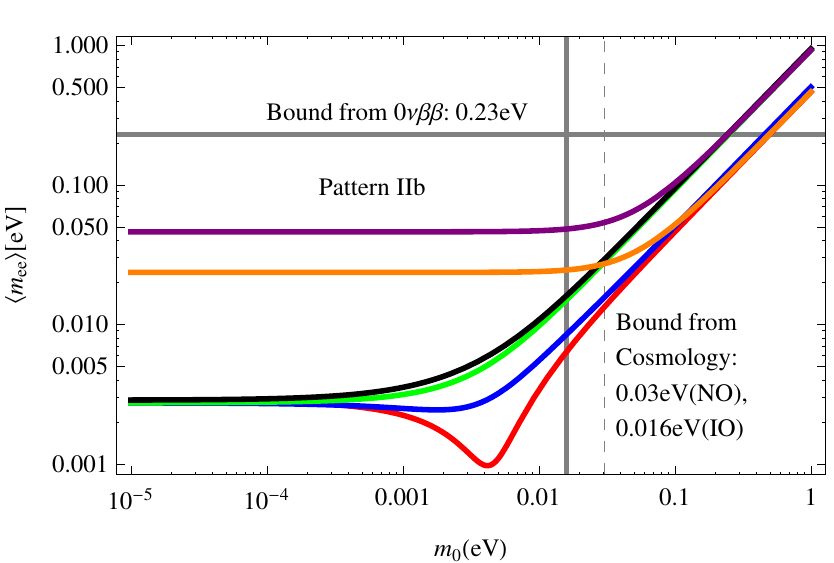}
\hfill
\caption{\label{Fig:2} The effective mass of neutrinoless double-beta decay $<m_{ee}>$ against the mass of the lightest
neutrino $m_{0}$ with the best fit data ($\theta^{bf}_{e}$,~$\theta^{bf}_{\nu}$). The bound on $m_{0}$ from cosmology is taken from the Ref.~\cite{74}. The constraint on $<m_{ee}>$ is from the Ref.~\cite{75}. The best fit data of $\Delta m^{2}_{12}$, $|\Delta m^{2}_{13}|$ is from the Ref.~\cite{65}. The legends for every pattern are shown in the top left panel. The indexes (j, k) are defined in Eq.~\ref{eq:23}. The best fit data ($\theta^{bf}_{e}$,~$\theta^{bf}_{\nu}$) for Pattern Ia and that for Pattern Ib take the second one in Tab. \ref{tab:2} respectively.}
\end{figure}

\begin{figure}[tbp]
\centering 
\includegraphics[width=0.48\textwidth]{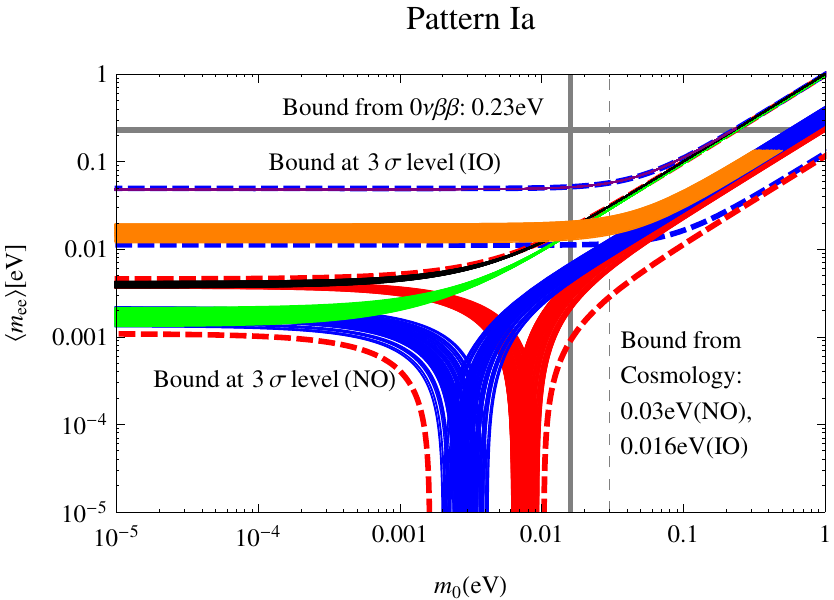}
\hfill
\includegraphics[width=0.48\textwidth]{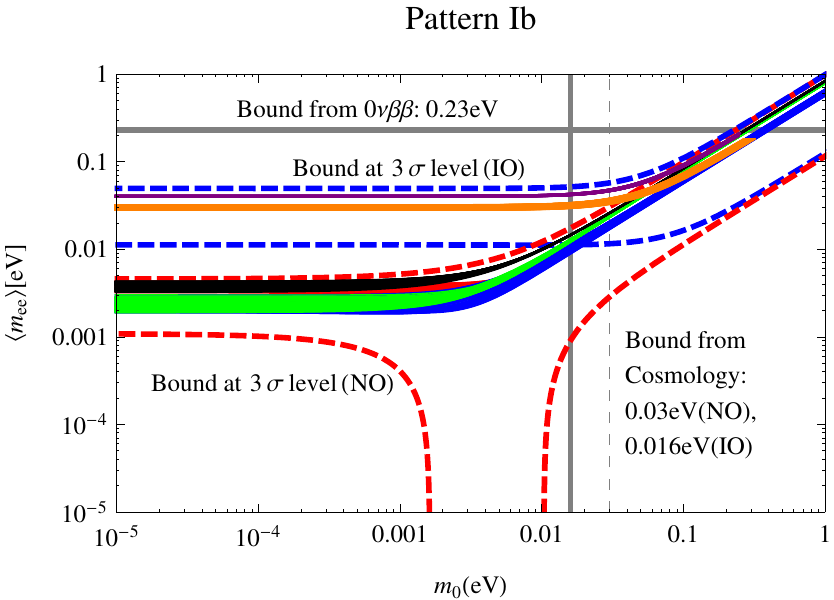}
\hfill
\includegraphics[width=0.48\textwidth]{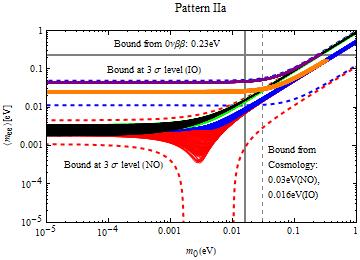}
\hfill
\includegraphics[width=0.48\textwidth]{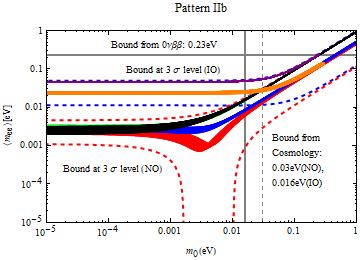}
\hfill
\caption{\label{Fig:3} The effective mass of neutrinoless double-beta decay $<m_{ee}>$ against the mass of the lightest neutrino $m_{0}$ in the $3\sigma$ ranges of ($\theta_{e}$,~$\theta_{\nu}$). The conventions follow those in Fig. \ref{Fig:2}. The dashed boundary lines at the $3\sigma$ level are obtained from the global fit data \cite{65}. }
\end{figure}

Although the residual symmetries ($Z_{2e}\times CP_{e}$, $Z_{2\nu}\times CP_{\nu}$) cannot restrain the masses of neutrinos $m_{ i}$ with $i=1, 2, 3$, they may affect the effective mass of neutrinoless double-beta decay $<m_{ee}>$ through the mixing angles and Majorana phases.
Here $<m_{ee}>$ is expressed as
\begin{equation}
\label{eq:77}
<m_{ee}>\equiv|m_{1}U^{2}_{11}+m_{2}U^{2}_{12}+m_{3}U^{2}_{13}|.
\end{equation}
Employing the lepton mixing matrix $U(\theta^{bf}_{e},~\theta^{bf}_{\nu}$) and the best fit data on $\Delta m^{2}_{12}$, $|\Delta m^{2}_{13}|$\cite{65}, we plot $<m_{ee}>$ against the mass of the lightest
neutrino $m_{0}$ in Fig.~\ref{Fig:2}. For $(\theta_{e},~\theta_{\nu}$) taken from the $3\sigma$ range around the best fit data, the curves of $<m_{ee}>$ in every pattern are shown in Fig.~\ref{Fig:3}. We make some
comments on the main results shown in these figures.\\
(i). In the case of IO, these patterns give stringent constraints on the ranges of $<m_{ee}>$. Specially, $<m_{ee}>$ for patterns with the indexes (1, 0) and (1, 1) is independent of the parameters $(\theta_{e},~\theta_{\nu}$). In the range of $m_{0}$ favored by cosmology, $<m_{ee}>$ is around 0.045eV for Pattern Ia, IIa, IIb. For Pattern Ib, it is 0.04eV. Theses values approximate the upper limit from the global fit data at the $3\sigma$ level. They are in the reach of future double beta decay experiments\cite{76}.  Interestingly, similar observations still hold for the patterns from the Group $S_{4}$ with GCP \cite{62}.\\
(ii). In the case of NO, the variance of $<m_{ee}>$ is noticeable for every pattern. Specially, for pattern Ia, Ib,  $<m_{ee}>$ with the indexes (1, 1) could reach the upper limit at the $3\sigma$ level. Even so, it is not accessible to near future experiments. \\
(iii). In both NO and IO case, without the precise constraint on the Dirac CP phase, these four mixing patterns cannot be discriminated by future double beta decay experiments because of the overlaps of ranges of $<m_{ee}>$.

\section{Summary}
For the group $PSL_2(7)$ with GCP symmetries, the predictions of the residual symmetries $Z_{2}\times CP$ in both neutrinos and charged leptons sector are examined.  Seven types of viable mixing patterns at the $3\sigma$ level of the global fit data are obtained. Among them, six types are paired through the complex conjugation. Three types of patterns can give the Dirac CP phase which is in the $1\sigma$ range of the global fit data.
With the parameters ($\theta_{e}$,~$\theta_{\nu}$), the constraints of residual symmetries on the effective mass of neutrinoless double-beta decay are also examined. In the case of IO, every pattern can give the effective mass accessible to the future experiments. \\

{\bf Competing Interests}\\
The author declares that there is no conflict of interest regarding the publication of this paper.

\acknowledgments
This work was supported by the National Natural Science Foundation of China under grant No.
11405101, 11705113, the Guangxi Scientific Programm Foundation under grant No. Guike AD19110045, and the research foundation of Gunlin University of Technology under
grant No. GUTQDJJ2018103.

\end{document}